\documentclass[
aps,%
11pt,%
final,%
notitlepage,%
oneside,%
onecolumn,%
nobibnotes,%
nofootinbib,%
superscriptaddress,%
showpacs,%
centertags]%
{revtex4}
\usepackage[cp1251]{inputenc}
\usepackage[english]{babel}
\usepackage{hhline}
\usepackage{amsmath,amssymb}
\usepackage{bm}
\usepackage{graphicx}
\begin {document}
\title{On the energy dependence of the muon transfer rate\\
from hydrogen to oxygen}
\author{\firstname{S.~V.}~\surname{Romanov}}
\email[]{Romanov_SVi@nrcki.ru}
\email[]{Serguei.V.Romanov@gmail.com}
\affiliation{National Research Centre "Kurchatov Institute",
Moscow, 123182, Russia.}
%
%
\begin{abstract}
The results of calculations of the muon transfer rate from the
$1s$ state of muonic hydrogen to the nucleus of a free oxygen atom
are presented in the interval of collision energies from $10^{-4}$
to 10~eV. The calculations were performed within a version of the
perturbed stationary states method proposed earlier. The electron
screening in the entrance channel of the transfer reaction was
taken into account. A $p$--wave resonance in the transfer rate is
predicted at collision energies of about 0.1~eV. This result is of
interest in the context of the planned laser experiment on precise
measurements of the hyperfine splitting energy of the $1s$ state
of muonic hydrogen.
\end{abstract}
\maketitle
\noindent{\it This is a preprint of the Work accepted for
publication in Physics of Atomic Nuclei,\\
$\copyright$ Pleiades Publishing, Ltd., 2022.}
\underline{http://pleiades.online/}
%
%
\section{\rm Introduction}
\label{intro}
%
%
The subject of the present study is the direct transfer of a
negative muon $\mu$ from the $1s$ state of a muonic hydrogen atom
to a nucleus of oxygen:
\begin{equation}
\label{react1}
\mu p\,(1s)+{\rm O}\to {\mu\,{\rm O}\,}^*+p\,.
\end{equation}
Here $p$ is a proton, ${\mu\,{\rm O}\,}^*$ is a muonic oxygen atom
in an excited state. Let $q(E)$ be the muon transfer rate reduced
by tradition to the atomic liquid--hydrogen density of $N_{\,\rm
H}=4.25\times 10^{\,22}\;\mbox{cm}^{-3}$:
\begin{equation}
\label{rate0}
q(E)=N_{\,\rm H}\,v\,\sigma(E)\,,
\end{equation}
$E$ is the collision energy, $v=\sqrt{2E/M}$ is the relative
velocity of the colliding particles, $M$ is their reduced mass,
$\sigma(E)$ is the total reaction cross section summed over all
final states of muonic oxygen. In an experimental study of
reactions of the type of (\ref{react1}) in dense gaseous mixtures,
the rate $\lambda(T)$ of the muon transfer from thermalized muonic
hydrogen atoms is measured. It depends on the mixture temperature
$T$ and, being reduced to the density $N_{\,\rm H}$\,, it is
obtained by averaging the rate $q(E)$ over the Maxwellian
distribution of relative velocities of the colliding particles.

The muonic hydrogen atom is an electrically neutral object whose
size is two orders of magnitude less than dimensions of ordinary
atoms. In this respect, it is similar to the neutron. The
reaction~(\ref{react1}) is exothermic with the excess kinetic
energy of a few keV. It is well known that at low collision
energies, when the $s$--wave contribution predominates, the cross
section of such a reaction is inversely proportional to the
relative velocity: $\sigma(E)\propto 1/v$ \cite{landau}. In this
case, the reaction rate $q(E)$ is independent of the collision
energy, and the rate $\lambda(T)$ is independent of the
temperature. The reaction~(\ref{react1}) was experimentally
studied in gaseous hydrogen--oxygen mixtures in Ref.~\cite{adam1}.
The mixtures were kept at room temperature and under pressure from
3 to 15 bar; the oxygen concentration was a few parts per
thousand. Delayed X--ray quanta emitted by muonic atoms of oxygen
were observed. The time dependence of the counting rate of these
quanta was found to be not purely exponential. This fact was
interpreted as a manifestation of the energy dependence of the
muon transfer rate at epithermal energies of $E=0.1-0.2$~eV. In
order to describe observed results quantitatively, the authors of
Ref.~\cite{adam1} performed a simulation of the slowing down of
muonic hydrogen atoms in gaseous mixtures with allowance for the
muon transfer to oxygen. The energy dependence of the muon tranfer
rate $q(E)$ was assumed to be a step function. The best fit to
experimental data was obtained with the following function:
\begin{equation}
\label{rate1}
q(E)=\left\{\begin{array}{rr}
8.5\,(2)\times 10^{\,10}\,{\mbox{ s}}^{-1}\,,&E<0.12\,\mbox{ eV}\,;\\
3.9\left(+0.5\atop -1.3\right)\times
10^{\,11}\,{\mbox{ s}}^{-1}\,,&
0.12<E<0.22\,\mbox{ eV}\,.\
\end{array}\right.
\end{equation}
The upper value is the rate of the muon transfer from muonic
hydrogen atoms thermalized at room temperature, the lower value
corresponds to epithermal energies. The epithermal transfer rate
is seen to be almost five times larger. In connection with this
result, the authors of Ref.~\cite{adam2} proposed to use the muon
transfer to oxygen in a laser experiment on precise measurements
of the hyperfine splitting energy of the $1s$ state of muonic
hydrogen. Let us consider the idea of this experiment.

Similarly to the ordinary hydrogen atom, the energy level of the
$1s$ state of muonic hydrogen is split into two components
specified by values of the quantum number $F$ of the total angular
momentum. Actually, $F$ is equal to 0 and 1, and the state with
$F=0$ has the lowest energy. The splitting energy $\Delta
E_{\,1s}$ is mainly determined by the first--order correction of
the perturbation theory in the interaction of the particle
magnetic moments. In the case of the ordinary hydrogen atom, the
respective result is well known~\cite{landau}. The replacement of
the magnetic moment of the electron by the one of the muon and
taking into account the reduced mass of muonic hydrogen yield
$\Delta E_{\,1s}\approx 0.18$~eV. The wavelength of the transition
between the hyperfine components is equal to $6.8~{\rm\mu m}$.
Concerning the lifetime of the $1s$ state, in hydrogen with a
small admixture of oxygen it is mainly determined by both the muon
decay and the muon transfer to oxygen. The muon decay rate is
equal to $4.5\times 10^{\,5}~\mbox{s}^{-1}$. Let us estimate the
muon transfer rate. In accordance with the result~(\ref{rate1}),
it is reasonable to set it equal to $10^{\,11}~\mbox{s}^{-1}$ at
the atomic liquid--hydrogen density. Then, for example, at the
pressure of 40 bar and the relative oxygen concentration of
$10^{-4}$, the muon transfer rate becomes comparable to the muon
decay rate, and the lifetime of the $1s$ state is about $1~{\rm\mu
s}$.

A muonic hydrogen atom in the $1s$ state is formed as a result of
the cascade capture of a muon. Both the components of the
hyperfine structure are populated in this process, and the kinetic
energy of the muonic atom can vary within wide limits. When
migrating through hydrogen gas, the muonic atom is slowed down by
losing its kinetic energy in collisions with hydrogen molecules.
Moreover, the interaction of the magnetic moments of the muon and
protons of molecules leads to the muon spin flip. As a result, the
muonic hygrogen atom proves to be in the lower
hyperfine--structure state with $F=0$. According to results of
simulations performed in Ref.~\cite{adam3}, at the pressure of 40
bar and the initial kinetic energy of muonic hydrogen of about
20~eV, the thermalization time is about 150~ns. The typical time
of the spin flip is an order of magnitude less (10--15~ns). Thus,
after thermalizing almost all muonic atoms are in the lower state
of the hyperfine structure. If now a laser pulse passes through
the gas, a part of muonic atoms will be found in an upper state
with $F=1$. Colliding with hydrogen molecules, these atoms go back
to the state with $F=0$. In each such collision, the transition
energy of 0.18~eV is divided between the muonic atom and the
hydrogen molecule. Taking into account their mass ratio and
rotational transitions in the molecule, it is possible to assert
that the muonic atom gets an additional kinetic energy of about
0.1~eV. Thus, after passing the laser pulse, the gas contains both
thermalized and epithermal muonic atoms in the state with $F=0$.
It is obvious that the dependence of the number of epithermal
muonic atoms on the laser radiation frequency has a resonant
character. If the shape of this dependence is known from an
experiment, the hyperfine splitting energy can be determined. If
the muon transfer reaction to a heavier nucleus is used to record
epithermal muonic atoms, the transfer from thermalized muonic
atoms is an unwanted background. The most suitable reaction is the
one whose rate increases sharply as the collision energy changes
from thermal to epithermal values. The muon transfer to oxygen
satisfies this requirement. In this case, the observable quantity
is the number of delayed X--ray quanta emitted by muonic atoms of
oxygen. Since this number is proportional to the amount of the
muonic atoms, it should also depend resonantly on the laser
radiation frequency.

In order to choose optimal conditions for the laser experiment,
new measurements of the rate of the reaction~(\ref{react1}) were
performed in Refs.~\cite{adam4,adam5}. The authors set a goal to
determine the energy dependence of the reaction rate at thermal
and epithermal energies. In order to avoid uncertainties related
to the initial energy distribution of muonic hydrogen atoms, the
muon transfer from thermalized muonic atoms was observed, and the
temperature dependence of the rate $\lambda(T)$ was studied. The
gaseous hydrogen--oxygen mixture used in this experiment was under
the pressure of 41~bar, the oxygen concentration was equal to
190~ppm, and the temperature varied from 104 to 300~K. Similarly
to Ref.~\cite{adam1}, time spectra of X--ray quanta radiated by
muonic atoms of oxygen were observed. The transfer rate was
determined from the slope of the part of these spectra that is
more than one microsecond apart from the instant of formation of
muonic oxigen atoms. The obtained results are presented in
Table~\ref{tab1}. They were first published in Ref.~\cite{adam4}
and then, after some corrections, in Ref.~\cite{adam5}. The new
values of the transfer rate obtained at room temperature agree
well with the result~(\ref{rate1}). One should pay attention to
the fast growth of the transfer rate with temperature. This fact
means that the $1/v$ law for the reaction cross section is not
valid at collision energies corresponding to the temperatures
considered. In order to reproduce the temperature dependence
obtained in the experiment, the authors of Ref.~\cite{adam4}
proposed a simple polynomial approximation of the energy
dependence of the transfer rate, which, in the authors' opinion,
is valid in the collision--energy interval of 0.01 -- 0.1~eV:
\begin{equation}
\label{rate2}
q(E)=p_{\,1}+p_{\,2}\,E+p_{\,3}\,E^{\,2}\,.
\end{equation}
The values of the coefficients $p_{\,i}$ are given in
Table~\ref{tab2}. They were obtained by averaging the
function~(\ref{rate2}) over the Maxwellian distribution and
fitting the result to the experimental data presented in
Table~\ref{tab1}. If the central values of the coefficients are
taken, the maximum of the function~(\ref{rate2}) is reached at the
energy of 0.097~eV. The maximum value of the transfer rate is
$1.74\times 10^{\,11}\;\mbox{s}^{-1}$, and it agrees qualitatively
with the result~(\ref{rate1}) for epithermal energies.
Measurements of the transfer rate $\lambda(T)$ were recently
extended towards lower and higher temperatures~\cite{adam6}. The
obtained results are also given in Table~\ref{tab1}. It is seen
that the transfer rate slowly decreases at $T\leq 104$~K. This
corresponds to the statement that the transfer rate becomes
temperature independent at low $T$. At $T>300$~K, the transfer
rate continues to increase.
%
%
%
%
\begin{table}[h]
\caption{Experimental values of the rate $\lambda(T)$ of the muon
transfer from thermalized muonic hydrogen atoms to oxygen at
various temperatures. The errors are given in the form
$\pm\;\sigma_1\pm\,\sigma_2$\,, where $\sigma_1$ includes
statistical and systematic errors related to the background
substraction from X--ray energy spectra, and $\sigma_2$ represents
other systematic errors. The value marked by an asterisk was
measured in the interval of 60 -- 79~K, and it was attributed to
the temperature of 70~K.}
\medskip
\begin{tabular}{|c|c|c|c|}
\hline
$\quad T,~{\rm K}\quad$&\multicolumn{3}{c|}{$\lambda(T),\,10^{\,10}\,\mbox{s}^{-1}$}\\
\cline{2-4}
   &\cite{adam4}&\cite{adam5}&\cite{adam6}\\
\hline
70&$-$&$-$&$\quad\hspace{-3mm}*\, 2.67\pm 0.40\pm 0.32\quad$\\
80&$-$&$-$&$\quad 2.96\pm 0.11\pm 0.36\quad$\\
104&$\quad 3.25\pm 0.10\pm 0.07\quad$&$\quad 3.07\pm 0.29\pm 0.07\quad$&$-$\\
153&$5.00\pm 0.11\pm 0.10$&$5.20\pm 0.33\pm 0.10$&$-$\\
201&$6.38\pm 0.10\pm 0.13$&$6.48\pm 0.32\pm 0.13$&$-$\\
240&$7.62\pm 0.12\pm 0.16$&$8.03\pm 0.35\pm 0.16$&$-$\\
272&$8.05\pm 0.12\pm 0.17$&$8.18\pm 0.37\pm 0.17$&$-$\\
300&$8.68\pm 0.12\pm 0.18$&$8.79\pm 0.39\pm 0.18$&$-$\\
323&$-$&$-$&$\quad 8.88\pm 0.62\pm 0.66\quad$\\
336&$-$&$-$&$\quad 9.37\pm 0.57\pm 0.70\quad$\\
\hline
\end{tabular}
\label{tab1}
\end{table}
%
%
%
%
\begin{table}[h]
\caption{Coefficients of the quadratic trinomial~(\ref{rate2})
approximating the energy dependence of the muon transsfer rate in
the collision--energy interval of 0.01 -- 0.1~eV~\cite{adam4}.}
\bigskip
\begin{tabular}{|c|c|c|}
\hline
$p_{\,1},\;\mbox{s}^{-1}$&$p_{\,2},\;\mbox{s}^{-1}\:\mbox{eV}^{-1}$&$p_{\,3},\:\mbox{s}^{-1}\;\mbox{eV}^{-2}$\\
\hline
$\quad (\,-1.32\pm 0.61\,)\times 10^{\,10}\quad$&$\quad (\,3.85\pm
0.54\,)\times 10^{\,12}\quad$&$\quad (\,-1.98\pm 0.65\,)\times 10^{\,13}\quad$\\
\hline
\end{tabular}
\label{tab2}
\end{table}
%
%
%
%
\begin{table}[h]
\caption{Experimental $\lambda_{\,\rm exp}$ and theoretical
$\lambda_{\,\rm th}$ values of the rate $\lambda(T)$ of the muon
transfer from thermalized muonic hydrogen atoms to oxygen at room
temperature. All the rates are given in units of
$10^{\,10}\,\mbox{s}^{-1}$. The rate obtained in Ref.~\cite{ger}
was reduced to the density $N_{\,\rm H}=4.25\times
10^{\,22}\;\mbox{\rm cm}^{-3}$. The value from Ref.~\cite{sul} is
the rate $q(E)$ at the mean thermal energy of 0.04~eV. The other
rates were obtained by averaging over the Maxwellian distribution.
Two values from Ref.~\cite{lin} correspond to the muon transfer to
a bare oxygen nucleus (the upper value) and to the nucleus of a
free atom with allowance for the electron screening (the lower
value).}
\smallskip
\begin{tabular}{|c|c|c|c|c|c|}
\hline
$\lambda_{\,\rm exp}$&\multicolumn{5}{c|}{$\lambda_{\,\rm th}$}\\
\cline{2-6}
\cite{adam1}           &      \cite{ger}&             \cite{sul}&      \cite{sav}&   \cite{fr1,fr2}&      \cite{lin}\\
\hline
$\quad 8.5\pm 0.2\quad$&$\quad 6.8\quad$&$\quad 7.7\pm 0.5\quad$&$\quad 8.4\quad$&$\quad 7.77\quad$&$\quad 23.2\quad$\\
                       &                &                       &                &                 &       4.42      \\
\hline
\end{tabular}
\label{tab3}
\end{table}
%
%
\newpage
Let us now consider results of available calculations of the muon
transfer rate~\cite{ger,sul,sav,fr1,fr2,lin}. It should be noted
that in the reaction~(\ref{react1}) the muon is transferred to a
nucleus of the oxygen molecule. In the calculations performed to
date, molecular effects were not taken into account, although they
may be significant at low collision energies. The calculations
were performed for the muon transfer either to a bare nucleus with
no electron shell or to the nucleus of a free atom. In the latter
case, the screening of the nuclear charge by atomic electrons was
considered in the polarization interaction between muonic hydrogen
and oxygen. The transfer rates calculated for muonic hydrogen
atoms thermalized at room temperature are presented in
Table~\ref{tab3}. All of them, with the exception of the results
of Ref.~\cite{lin}, are in reasonable agreement with experimental
data. The situation with the energy dependence of the transfer
rate is more complicated. In Ref.~\cite{ger}, the muon transfer to
a bare oxygen nucleus was considered within the Landau--Zener
model. In taking into account only the $s$--wave contribution, it
was found that, as the collision energy increases, the transfer
rate was initially constant and then began to decrease. A similar
dependence was obtained in Ref.~\cite{sul}. The respective
calculation was performed on the basis of Faddeev equations for
collision energies below 0.5~eV. As well as in Ref.~\cite{ger},
the muon transfer to a bare oxygen nucleus was considered, and
only the $s$--wave contribution was taken into account. An
important step was made in Ref.~\cite{sav}. The authors considered
contributions of partial waves with nonzero values of the orbital
angular momentum. They used the two--state approximation and the
Landau--Zener model in a more refined version than in
Ref.~\cite{ger}. Moreover, the electron screening in the entrance
channel of the transfer reaction was taken into account. As a
result, the existence of a $d$--wave resonance in the transfer
rate was predicted at collision energies of about 0.19~eV. The
resonance width and the peak value of the transfer rate agree
qualitatively with the experimental result~(\ref{rate1}) for
epithermal energies. Subsequently, the role of partial waves with
nonzero orbital angular momenta was studied in
Refs.~\cite{fr1,fr2}. The calculation of the transfer rate was
performed within the method of hyperspherical elliptic
coordinates; the electron screening was not considered. The
results of this calculation differ drastically from the ones of
Ref.~\cite{sav}. It was obtained that, in the energy interval
between $2\times 10^{\,-3}$ and 1~eV, the $p$--wave made the main
contribution to the muon transfer rate. As the collision energy
increases, the rate passes through a broad maximum at thermal
energies, reaching a value of about $8\times
10^{\,10}\,{\mbox{s}}^{-1}$, and then it decreases monotonically
up to the energy of 2~eV. Thus, the calculation performed in
Refs.~\cite{fr1,fr2} correctly reproduces the value of the
transfer rate from thermalized muonic atoms, but its results
contradict the experimental fact of the increase of the transfer
rate at epithermal energies. At last, one more calculation was
performed in Ref.~\cite{lin}. The authors used a version of the
hyperspherical functions method and considered the muon transfer
both to a bare oxygen nucleus and to the nucleus of a free atom
with allowance for the electron screening. In both the cases, a
broad $p$--wave maximum in the energy dependence of the transfer
rate was predicted, but its position and height proved to be very
sensitive to the electron screening. In the case of the transfer
to a bare nucleus, the maximum was also located at thermal
energies, but its height was about eight times larger than that
obtained in Refs.~\cite{fr1,fr2}. In taking into account the
electron screening, the curve of the energy dependence went down
strongly at energies below 0.1~eV. As a result, the maximum
shifted to the energy of 0.11~eV and became less sharp. The
maximum value of the transfer rate was obtained to be nearly equal
to $1.3\times 10^{\,11}\,{\mbox{s}}^{-1}$. This result agrees
qualitatively with the observed increase of the transfer rate at
epithermal energies. Concerning the transfer rate from thermalized
muonic atoms, the calculation did not reproduce its experimental
value closely. The rate obtained for the transfer to a bare
nucleus was about three times large. In this case, the electron
screening proved to be even more significant. It reduced the
transfer rate nearly by a factor of six down to a half of the
experimental value. Thus, the results of the calculations
performed in Refs.~\cite{sav,fr1,fr2,lin} show the importance of
taking into account partial waves with nonzero orbital angular
momenta and the electron screening. However, the energy
dependences of the transfer rate predicted in these studies are
substantially different.

In connection with the ambiguity of the theoretical results
considered above, in the present study, the rate of the
reaction~(\ref{react1}) was calculated for collision energies from
$10^{-4}$ to 10~eV. A version of the perturbed stationary states
method was used. It was proposed in Ref.~\cite{my1} and applied to
calculations of the muon transfer rate from hydrogen to neon in
Ref.~\cite{my2}. This method is based on a substantial difference
of the energies of relative motion in the reaction channels. The
reaction of the muon transfer to oxygen satisfies this condition.
The analysis of energy spectra of delayed X--ray quanta performed
in Ref.~\cite{adam1} showed that, in the reaction~(\ref{react1}),
muons were transferred to muonic oxygen states with the principle
quantum number $n\leq 6$. In this case, the kinetic energy of
relative motion of the reaction products is at least 2.4~keV, and
it is much greater than the above--mentioned collision energies in
the entrance channel. It is obvious that, in this case, it is
first of all necessary to provide an asymptotically correct
description of the entrance channel. Therefore, the wave function
of the three--body system (muon, proton, and oxygen nucleus) was
constructed in the form of an expansion in eigenfunctions of a
two--center Coulomb problem formulated in terms of the Jacobi
coordinates of the entrance channel. Since this method was
decribed in detail in Ref.~\cite{my1}, only its brief overview is
given in Sec.~\ref{method}. Some details of the calculation are
discussed in Sec.~\ref{details}. The results and conclusions are
presented in Sec.~\ref{results}. Unless otherwise stated, the
muon--atom units will henceforth be used:
\begin{equation}
\label{mau}
\hbar=e=m_\mu=1\,,
\end{equation}
$e$ is the proton charge, $m_\mu$ is the muon mass; the unit of
length is $2.56\times10^{\,-11}$~cm, and the unit of energy is
5.63~keV.
%
%
\section{\rm Method of calculations}
\label{method}
%
%
The reaction of the direct muon transfer to oxygen is a particular
case of the charge--exchange reaction in a collision between a
muonic atom of a hydrogen isotope $\rm H$ in the $1s$ state and a
nucleus of an atomic number $Z>1$:
\begin{equation}
\label{react2}
\mu{\rm H}\,(1s)+Z\to {\mu Z\,}^*+{\rm H}\,,
\end{equation}
where ${\mu Z\,}^*$ is a muonic atom of the element $Z$ in an
excited state. Let us introduce the Jacobi coordinates of the
entrance channel of this reaction: the vector $\bf r$ specifying
the muon position with respect to the hydrogen nucleus and the
vector $\bf R$ connecting the center of mass $C_2$ of the muonic
hydrogen atom with the $Z$ nucleus (Fig.~1). The center of mass
$C_3$ of the three--body system lies on the vector $\bf R$. Let us
introduce a few more quantities: the vector $\bf r_1$ drawn from
the point $C_2$ to the muon, the distance $r_2$ between the muon
and the $Z$ nucleus, and the distance $R_{\,{\rm H}Z}$ between the
nuclei. The nonrelativistic three--body Hamiltonian written in the
center--of--mass frame is:
\begin{equation}
\label{Htot}
\hat H=-\frac{1}{2M}\,{\Delta}_{\bf R}+{\hat H}_\mu
+\frac{Z}{R_{\,{\rm H}Z}}\,.
\end{equation}
The first term is the kinetic--energy operator of the relative
motion of the muonic hydrogen atom and the $Z$ nucleus, $M$ is the
reduced mass of the $Z$ nucleus with respect to the muonic atom:
\begin{equation}
\label{Mr}
M^{-1}=(M_{\rm H}+1)^{-1}+M_Z^{-1}\,,
\end{equation}
where $M_{\rm H}$ and $M_Z$ are the nuclear masses. The term
${\hat H}_\mu$ is the Hamiltonian of the muonic hydrogen atom
supplemented with the potential energy of the Coulomb interaction
of the muon and the $Z$ nucleus:
\begin{equation}
\label{Hmu1}
{\hat H}_\mu=-\frac{1}{2m_{\mu\rm H}}\,{\Delta}_{\,\bf r}
-\frac{1}{r}-\frac{Z}{r_2}\,,
\end{equation}
where $m_{\mu\rm H}$ is the reduced mass of the muonic hydrogen
atom:
\begin{equation}
\label{mrH}
m_{\mu\rm H}^{-1}=M_{\rm H}^{-1}+1\,.
\end{equation}
The last term in~(\ref{Htot}) describes the Coulomb repulsion of
the $\rm H$ and $Z$ nuclei.
\newpage
%
%
%
%
\begin{figure}[h]
\includegraphics[scale=0.8]{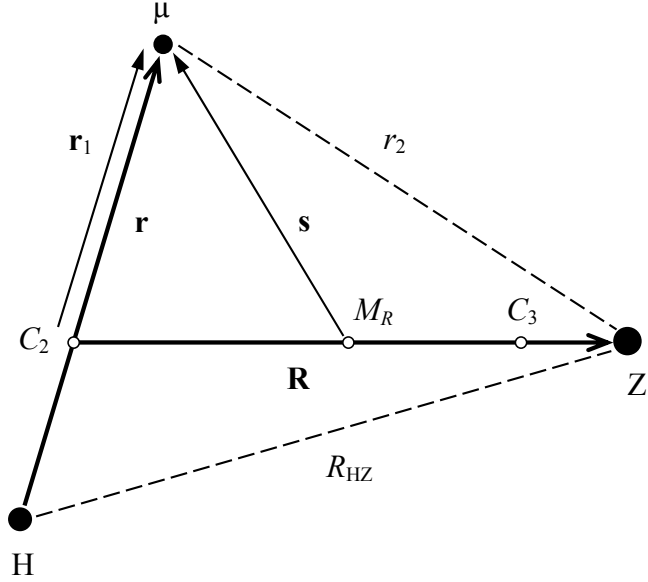}
\caption{Jacobi coordinates of the entrance channel of the muon
transfer reaction. $C_2$ is the center of mass of the muonic
hydrogen atom, $C_3$ is the center of mass of the three--body
system, $M_R$ is the midpoint of the vector $\bf R$.}
\end{figure}
%

Let us single out a two--center problem in the three--body
Hamiltonian. For this purpose, let us rewrite the term ${\hat
H}_\mu$ as follows~\cite{jap}:
\begin{equation}
\label{Hmu2}
{\hat H}_\mu=m_{\mu\rm H}\cdot{\hat h}_\mu\,,
\end{equation}
\begin{equation}
\label{Hmu3}
{\hat h}_\mu=-\frac{1}{2}\,{\Delta}_{\,\bf s} -\frac{1}{r_1}
-\frac{Z\,'}{r_2}\,.
\end{equation}
The vector $\bf s$ connects the midpoint $M_R$ of the vector $\bf
R$ with the muon:
\begin{equation}
\label{s}
{\bf s}={\bf r_1}-\frac{\bf R}{2}\,,
\quad
{\bf r_1}=m_{\mu\rm H}\cdot{\bf r}\,.
\end{equation}
The charge $Z\,'$ is:
\begin{equation}
\label{Z'def}
Z\,'=\frac{Z}{m_{\mu\rm H}}\,.
\end{equation}
The operator ${\hat h}_\mu$ is the Hamiltonian of a muon in the
field of two immovable Coulomb centers whose charges are equal to
unity and $Z\,'$. The unit charge is placed in the center of mass
of the muonic hydrogen atom, while the position of the charge
$Z\,'$ coincides with the one of the $Z$ nucleus. For the muon
transfer from protium to oxygen, the charge $Z\,'$ is:
\begin{equation}
\label{Z'oxygen}
m_{\mu\rm H}=0.899\,,
\quad
Z\,'=8.90\,.
\end{equation}

In the coordinate frame with the origin in the point $M_R$ and the
polar axis directed along the vector $\bf R$, the muon position is
specified by the prolate spheroidal coordinates $\xi$, $\eta$, and
$\varphi$~\cite{ponom}:
\begin{equation}
\label{psc}
\xi=\frac{r_1+r_2}{R}\,,
\quad
\eta=\frac{r_1-r_2}{R}\,.
\end{equation}
The azimuthal angle $\varphi$ lies in the plane that is orthogonal
to vector $\bf R$ and which contains the point $M_R$. Surfaces of
constant values of the coordinates $\xi$ and $\eta$ are
respectively prolate ellipsoids of revolution and two--sheeted
hyperboloids. These surfaces have the foci at the points $C_2$ and
$Z$ for which $\xi=1$ and $\eta=\mp 1$.

Let us consider the eigenvalue problem for the two--center
Hamiltonian ${\hat h}_\mu$:
\begin{equation}
\label{2cent}
{\hat h}_\mu\,\psi_{jm}(\xi,\eta;R)\,\frac{\exp (\pm im\varphi)}
{\sqrt{2\pi}}={\varepsilon_{jm}}(R)\,
\psi_{jm}(\xi,\eta;R)\,\frac{\exp (\pm im\varphi)}{\sqrt{2\pi}}
\,.
\end{equation}
Here the dependence on the angle $\varphi$ is explicitly singled
out, the quantum number $m$ takes nonnegative integer values, and
the index $j$ stands for the set of the remaining quantum numbers.
For bound states, these are either the numbers $n_\xi$ and
$n_\eta$ of nodes in the respective variables or the parabolic
quantum numbers $n_1$ and $n_2$ in the limit of
$R\to\infty$~\cite{ponom}. The two--center problem~(\ref{2cent})
is solved at a fixed distance $R$, which appears in eigenfunctions
and eigenvalues ${\varepsilon_{jm}}(R)$ as a parameter. The
functions $\psi_{jm}(\xi,\eta;R)$ can be chosen to be
real--valued. The eigenfunctions with identical $m$ and different
sets $i$ and $j$ of the remaining quantum numbers are
orthonormalized:
\begin{gather}
\label{orto}
\int\psi_{im}(\xi,\eta;R)\,\psi_{jm}(\xi,\eta;R)\,d\tau=
\delta_{ij}\,,\\
d\tau=(R/2)^3\,({\xi}^2-{\eta}^2)\,d\xi\,d\eta\,.
\end{gather}
Here the integral is taken over the region specified by the
inequalities $1\le\xi<\infty$ and $-1\le\eta\le+1$. The
orthonormality in $m$ is ensured by the factors $\exp(\pm
im\varphi)/\sqrt{2\pi}$. It is obvious that the solutions of the
problem~(\ref{2cent}) are also the eigenfunctions of the
Hamiltonian ${\hat H}_\mu$ with the eigenvalues $m_{\mu\rm
H}\cdot\varepsilon_{jm}(R)$.

It is well known that the two--center problem admits the
separation of variables in prolate spheroidal
coordinates~\cite{ponom}. The function $\psi_{jm}(\xi,\eta;R)$ is
the product of a radial and an angular function. The former
depends only on $\xi$, while the latter depends only on $\eta$. By
solving two differential equations for these functions with
respective boundary conditions, one can find the eigenvalue
${\varepsilon_{jm}}(R)$, the separation constant, and eventually
construct the eigenfunction. In the previous calculation of the
muon transfer rate to neon~\cite{my2}, this scheme was implemented
on the basis of standard expansions of the radial and angular
functions in infinite series~\cite{jaffe,baber,bates}. The
eigenvalue ${\varepsilon_{jm}}(R)$ and the separation constant
were determined on the basis of the method proposed in
Ref.~\cite{had} and modified for the case in which the expansion
coefficients are not monotonic functions of their number. This
scheme was also used in the present study.

The primary coordinate frame in which the motion of three
particles is considered has fixed axes and the origin at the
center of mass $C_3$. The Hamiltonian $\hat H$ commutes with the
operator ${\hat{\bf J}}^{\,2}$ of the square of the orbital
angular momentum of the three--body system and with the operator
${\hat J}_z$ of its projection onto the $z$--axis of the primary
coordinate frame. In addition, $\hat H$ commutes with the operator
$\hat{\rm P}$ of inversion of the spatial coordinates of all the
particles. It is convenient to take the eigenfunctions of these
three operators as basis solutions in which the three-body
wavefunction is expanded. Let us also require them to be solutions
of the two--center problem~(\ref{2cent}). Since, in this problem,
the muon coordinates are associated with the vector $\bf R$, let
us introduce the polar and azimuthal angles $\Theta$ and $\Phi$
specifying the orientation of $\bf R$ with respect to the axes of
the primary coordinate frame. Then a configuration of the
three-body system is specified by the six independent coordinates
$R$, $\Theta$, $\Phi$, $\xi$, $\eta$, and $\varphi$, and the basis
solutions are:
\begin{equation}
\label{basis}
\Psi_{Mjm}^{JP}(R,\Theta,\Phi,\xi,\eta,\varphi)=
\frac{{\chi}_{jm}^{JP}(R)}{R}\,\Upsilon_{Mm}^{JP}(\Phi,\Theta,
\varphi)\,\psi_{jm}(\xi,\eta;R)\,.
\end{equation}
Here ${\chi}_{jm}^{JP}(R)$ is a radial function that depends on
the quantum numbers indicated in the indices,
$\Upsilon_{Mm}^{JP}(\Phi,\Theta,\varphi)$ is the eigenfunction of
the operators ${\hat{\bf J}}^{\,2}$, ${\hat J}_z$, and $\hat{\rm
P}$ for the eigenvalues $J(J+1)$, $M$, and $P$. The nonnegative
integer $m$ introduced in Eq.~(\ref{2cent}) is the modulus of the
projection of the total orbital angular momentum of the
three--body system onto the direction of the vector $\bf R$. The
functions $\Upsilon_{Mm}^{JP}$ are orthonormalized:
\begin{equation}
\label{ornY}
\int\limits_0^{\pi}\sin\Theta\,d\Theta
\int\limits_0^{2\pi}d\Phi\int\limits_0^{2\pi}d\varphi\,
\bigl(\Upsilon_{Mm}^{JP}\bigr)^*\,\Upsilon_{M'm'}^{J'P'}=
\delta_{JJ'}\,\delta_{PP'}\,\delta_{MM'}\,\delta_{mm'}.
\end{equation}
Their specific form depemds on $m$. If $m=0$, then
\begin{equation}
\label{Y_M0}
\Upsilon_{Mm=0}^{JP}(\Phi,\Theta,\varphi)=
\frac{Y_{JM}(\Theta,\Phi)}{\sqrt{2\pi}}\,,
\end{equation}
where $Y_{JM}(\Theta,\Phi)$ is the spherical harmonic. In this
case, the parity is unambiguously determined by the quantum number
$J$: $P=(-1)^J$. If $m\neq 0$, then
\begin{equation}
\label{Y_Mm}
\Upsilon_{Mm}^{JP}(\Phi,\Theta,\varphi)=
\frac{\sqrt{2J+1}}{4\pi}\left[\,
(-1)^mD_{Mm}^{\,J}(\Phi,\Theta,\varphi)+
P\,(-1)^JD_{M(-m)}^{\,J}(\Phi,\Theta,\varphi)\,\right],
\end{equation}
where $D_{Mm}^{\,J}$ and $D_{M(-m)}^{\,J}$ are the Wigner
functions~\cite{dav} transformed under the inversion as follows:
\begin{equation}
\label{InvD}
D_{Mm}^{\,J}(\Phi,\Theta,\varphi)\longrightarrow
(-1)^{J-m}D_{M(-m)}^{\,J}(\Phi,\Theta,\varphi)\,.
\end{equation}
In this case, two parity values are possible at given $J$: $P=\pm
(-1)^J$.

Let us consider the time--independent Schr{\"o}dinger equation for
the three-body wavefunction with the quantum numbers $J$, $M$, and
$P$:
\begin{equation}
\label{Schr}
\hat H\,\Psi_{M}^{JP}=\tilde E\,\Psi_{M}^{JP}\,.
\end{equation}
For the reaction~(\ref{react2}), the energy of the system $\tilde
E$ is:
\begin{equation}
\label{Etot}
\tilde E=E_{\mu\rm H}(1s)+E\,,
\end{equation}
where $E_{\mu\rm H}(1s)$ is the ground--state energy of the muonic
hydrogen atom:
\begin{equation}
\label{EmuH}
E_{\mu\rm H}(1s)=-\frac{m_{\mu\rm H}}{2}\,,
\end{equation}
and $E$ is the collision energy:
\begin{equation}
\label{Ec}
E=\frac{M v^{\,2}}{2}=\frac{k^{\,2}}{2M}\,.
\end{equation}
Here $v$ is the velocity of the relative motion of the $\mu\rm H$
atom and the $Z$ nucleus at infinite separation, and $k=M v$ is
the asymptotic momentum of the relative motion.

Let us seek a solution of Eq.~(\ref{Schr}) in the form of an
expansion in the basis solutions~(\ref{basis}):
\begin{equation}
\label{expan}
\Psi_M^{JP}=\sum_{jm}\,\Psi_{Mjm}^{JP}\,.
\end{equation}
The substitution of this expansion into Eq.~(\ref{Schr}) and the
integration over the variables $\Theta$, $\Phi$, $\xi$, $\eta$,
and $\varphi$ with allowance for the orthonormality of the basis
solutions lead to a set of coupled second--order differential
equations for the radial functions ${\chi}_{jm}^{JP}(R)$. These
equations were presented in Ref.~\cite{my1}. In practice, the
contribution of a finite number of two--center states is taken
into account in the expansion~(\ref{expan}). After solving the
obtained set of equations with respective boundary conditions, one
can calculate the total cross section of the
reaction~(\ref{react2}).

As already noted, the main idea of the method under discussion
consists in constructing an asymptotically correct description of
the entrance reaction channel at large values of the distance $R$.
In the limit of $R\to\infty$, the solutions of the two--center
problem~(\ref{2cent}) split into two groups. The states of one
group are localized near the left center, which is placed in the
center of mass of the $\mu\rm H$ atom and carries the unit charge.
The states of the other group are localized near the right center
$Z\,'$. The simplest way to describe the entrance channel is to
take into account the only state of the left--center group. Its
asymptotic quantum numbers are:
\begin{equation}
\label{1sH}
m=n_1=n_2=0\,,\quad n=1\,,
\end{equation}
where $n_1$ and $n_2$ are the parabolic quantum numbers, and
$n=n_1+n_2+m+1$ is the principle quantum number. Hereinafter all
the quantities related to this state will be marked with the index
0. In the limit considered, the eigenfunction $\psi_0$ and the
eigenvalue $\varepsilon_0(R)$ of the two--center problem are:
\begin{equation}
\label{H1s}
\psi_0\propto\exp\,(-m_{\mu\rm H}\cdot r)\,,\quad
\varepsilon_0(R\to\infty)=-\frac{1}{2}\,.
\end{equation}
Thus, the two--center eigenfunction goes into the wave function of
the ground state of the $\mu\rm H$ atom with the correct value of
the reduced mass. This is because the left center lies at the
center of mass of the muonic hydrogen atom; the argument of the
exponent in the function $\psi_0$ is the distance from this center
to the muon. The eigenvalue of the Hamiltonian ${\hat H}_\mu$
tends to the correct dissociation limit:
\begin{equation}
\label{dis0}
m_{\mu\rm H}\cdot\varepsilon_0(R\to\infty)=E_{\mu\rm H}(1s)\,.
\end{equation}

At large $R$, the relative motion in the entrance channel is
determined by the potential $U_0(R)$, which is obtained by
averaging the three--body Hamiltonian over the state $\psi_0$. The
expansion of this potential in powers of $R^{\,-1}$ was considered
in Ref.~\cite{my1}. The leading term of this expansion is
proportional to $R^{\,-4}$ and corresponds to the polarization
attraction between the muonic hydrogen atom and the $Z$ nucleus:
\begin{equation}
\label{U0}
U_0(R)=-\frac{\beta_0 Z^{\,2}}{2R^{\,4}}\,.
\end{equation}
The polarizability of the muonic atom was found to be:
\begin{equation}
\label{beta0}
\beta_0=\beta\left[\,1-\frac{1}{(M_{\rm H}+1)^{\,2}}\,\right]\,,
\end{equation}
where $\beta$ is the exact value of the polarizability:
\begin{equation}
\label{beta1}
\beta=\frac{9}{2\,m_{\mu\rm H}^{\,3}}\,.
\end{equation}
It should be noted that, because of the cube of the reduced mass
in the denominator of this expression, the value of $\beta$ can
differ markedly from the frequently used value of 4.5\,, which
corresponds to an infinitely heavy nucleus $\rm H$. In particular,
$\beta\approx 6.2$ for muonic protium. Although $\beta_0$ does not
coincide with $\beta$, their values are very close. For muonic
protium, $\beta_0\approx 0.99\,\beta$\,. The distinction between
these values is due to the fact that the Coulomb repulsion of the
nuclei (the last term in Eq.~(\ref{Htot})) is not diagonal in the
two-center basis. As was shown in Ref.~\cite{my1}, taking this
circumstance into account leads to a small correction to the
polarizability. Its addition to $\beta_0$ yields $\beta$ exactly.
Thus, the use of only one state of the left center already
provides a good description of the entrance reaction channel at
large $R$: the dissociation limit is correct, no spurious
long--range interactions appear (at least to terms of order
$R^{-4}$ inclusive), and the polarizability of the muonic hydrogen
atom is reproduced to within $1~\%$. Therefore, this description
will be used in the following. Moreover, since the values of
$\beta$ and $\beta_0$ are close, the polarization potential with
the exact value of $\beta$ will be used to describe the relative
motion in the entrance channel at large $R$:
\begin{equation}
\label{Up}
U_p(R)=-\frac{\beta Z^{\,2}}{2R^{\,4}}\,.
\end{equation}

Within the approach considered, the muon transfer channel is
described by states of the right center. In the limit of
$R\to\infty$, they correspond to the $\mu Z\,'$ atom with an
infinitely heavy nucleus rather than to the real $\mu Z$ atom. In
particular, the wave functions of these states do not contain the
reduced mass of the $\mu Z$ atom at all. Moreover, the equations
for the radial functions of the relative motion in the transfer
channel remain coupled even at an infinitely large distance $R$.
The reason of these difficulties is the use of the
entrance--channel Jacobi coordinates, which are not natural for
the transfer channel. It is obvious that, in this case, it is
impossible to calculate cross sections of the muon transfer to
individual states of the $\mu Z$ atom. Nevertheless, since the
states of the right center are asymptotically localized near the
$Z$ nucleus, a group of these states as a whole describes the
migration of the muon charge cloud from $\rm H$ to $Z$, i.e. the
muon transfer. Therefore, it is possible to calculate the total
transfer cross section. This calculation is based on the fact
that, at large $R$, the basis two--center functions describing the
entrance and transfer channels are localized at the different
centers. Therefore, as $R$ increases, the matrix elements of the
three--body Hamiltonian that couple the equations for the radial
functions of these channels decrease exponentially. As a result,
at $R\to\infty$, the set of radial equations splits into two
groups, which describe the channels separately. In the simplest
approximation in which the only left--center state with the
quantum numbers~(\ref{1sH}) is taken into account, the entrance
channel is asymptotically described by one equation:
\begin{equation}
\label{xi0}
\frac{d^{\,2}{\chi}_0^J}{dR^{\,2}}+
\left[\,k^{\,2}-\frac{J(J+1)}{R^{\,2}}
-2MU_p(R)\,\right]\chi_0^J=0\,,
\end{equation}
where ${\chi}_0^J$ is the radial function of the entrance channel.
Its upper index $P$ is omitted because at $m=0$ the parity is
unambiguously determined by the quantum number $J$: $P=(-1)^J$.
The boundary condition for this function at large $R$ is:
\begin{equation}
\label{xi0as}
{\chi}_0^J\,(\,R\to\infty\,)\longrightarrow\sin\,(\,kR-J\pi/2\,)+Q_0^{\,J}
\exp\left[\,i\,(\,kR-J\pi/2\,)\,\right]\,.
\end{equation}
The complex amplitude $Q_0^{\,J}$ depends on $J$ and $k$. The
radial functions of the transfer channel behave asymptotically as
divergent waves. A method of constructing such solutions for
coupled equations was described in Ref.~\cite{my1}. The boundary
condition at $R=0$ is standard: the radial functions vanish at
this point.

By integrating the set of coupled radial equations under the above
boundary conditions, it is possible to construct the amplitudes
$Q_0^{\,J}$. Let us rewrite the asymptotic radial function
${\chi}_0^J$ in the following form:
\begin{equation}
\label{xi0as1}
{\chi}_0^J\,(\,R\to\infty\,)\propto
\exp\left[\,-i\,(\,kR-J\pi/2\,)\,\right]
-S_0^J\exp\left[\,i\,(\,kR-J\pi/2\,)\,\right]\,,
\end{equation}
where $S_0^J$ is the diagonal $S$--matrix element corresponding to
the entrance channel:
\begin{equation}
\label{S0J}
S_0^J=1+2\,i\,Q_0^{\,J}\,.
\end{equation}
Since the muon transfer is the only inelastic channel at the
collision energies considered, the total cross section of the muon
transfer is~\cite{dav}:
\begin{equation}
\label{crsec}
\sigma(E)=\frac{\pi}{k^{\,2}}\sum_{J=0}^{\infty}\,(\,2J+1\,)
\left(1-{|\,S_0^J\,|}^{\,2}\,\right)\,.
\end{equation}
The muon transfer rate $q(E)$ considered as a function of the
collision energy and normalized to the atomic density of liquid
hydrogen is calculated by the formula~(\ref{rate0}). The rate
$\lambda(T)$ of the muon transfer from thermalized muonic hydrogen
atoms is obtained by averaging the rate $q(E)$ over the Maxwellian
distribution of relative velocities in the entrance channel.
%
%
\section{\rm Some details of the calculation}
\label{details}
%
%
Let us consider how the method described in Sec.~\ref{method} is
applied to the calculation of the rate of the muon transfer from
hydrogen to oxygen. At large interatomic distances, the entrance
reaction channel is described by one left--center state $\psi_0$
with the quantum numbers~(\ref{1sH}). In order to choose the
relevant right--center states, let us take advantage of the
generally accepted viewpoint that the muon transfer from hydrogen
to a heavier nucleus is mainly due to quasicrossings of adiabatic
terms associated with the reaction channels. The right--center
states will be specified by the parabolic quantum numbers $n'_1$
and $n'_2$\,, and by the principle quantum number
$n'=n'_1+n'_2+m+1$. As is known~\cite{ponom}, quasicrossings are
possible only for terms with identical values of the numbers $m$
and $n_1$. Since these numbers are equal to zero for the state
$\psi_0$, let us consider right--center states with $m=n'_1=0$.
Their wave functions have no nodes in the variable $\xi$ but
differ in the number $n_\eta$ of nodes in the variable $\eta$. In
the following, the states with $3\leq n'_2\leq 6$ will be of
interest. According to the relation between $n_\eta$ and
$n'_2$~\cite{ponom}, $n_\eta=n'_2$ for these states at
$Z\,'=8.90$. The dependences of the eigenvalues ${\varepsilon_j}$
of the two--center problem~(\ref{2cent}) on the interatomic
distance $R$ are shown in Fig.~2 (the index $j$ is now reduced to
the numbers $n'_1$ and $n'_2$\,). Quasicrossings occur in the
following regions of $R$: $4-6$, $7-8$, and $12-13$. If we begin
from the $n'_2=3$ term and move toward large $R$, then each
transition from one term to another in a quasicrossing region
increases $n'_2$ by unity. This corresponds to the general
rule~\cite{ponom} that the terms involved into a quasicrossing
differ in $n_\eta$ by unity. There are two more quasicrossings not
shown in Fig.~2. One of them lies at $R\approx 24.3$. For it, the
number $n'_2$ increases from 6 to 7. Finally, the farthest
quasicrossing lies at $R\approx 66$. It involves the right--center
state with $n'_2=7$ and the state $\psi_0$ for which the number of
nodes $n_\eta=8$.

Let us assume that the muonic hydrogen atom is at a very large
distance $R$ from the oxygen nucleus. In this case, the function
$\psi_0$ is localized in a vicinity of the proton, where it
coincides nearly with the wave function of the ground state of an
isolated muonic hydrogen atom. All eight nodes of this function
lie in a vicinity of the oxygen nucleus, where $\psi_0$ is
exponentially small. As $R$ decreases, this pattern remains
unchanged up to the first quasicrossing with the right--center
state with $n'_2=7$. After passing through the quasicrossing
region, the muon charge distribution in these states changes
abruptly. In the $\psi_0$ state, the muon charge cloud migrates to
the oxygen nucleus and becomes exponentially small near the
proton. In the $n'_2=7$ state, the charge, on the contrary, flows
to the proton, and all seven nodes of the wave function prove to
be in a vicinity of the oxygen nucleus, where the wave function is
exponentially small. A similar picture is observed in passing
through other quasicrossings, which occur deep under the potential
barrier separating the Coulomb wells of the two--center problem.
In this case, it can be argued that, between narrow quasicrossing
regions, the muon charge cloud is localized near one of the
Coulomb centers and, as $R$ decreases, the right-center states
with the number $n'_2$ successively decreasing by unity describe
the muonic hydrogen atom in the Coulomb field of the oxygen
nucleus. In particular, this is the $n'_2=6$ state in the region
of $13<R<24$. The fact that it does correspond to the muonic
hydrogen atom in the field of the oxygen nucleus is confirmed by
calculations of the adiabatic potential that is equal to the sum
of the eigenvalue \mbox{$m_{\mu\rm H}\cdot\varepsilon_j (R)$} of
the Hamiltonian ${\hat H}_\mu$ counted from the energy $E_{\mu\rm
H}(1s)$ of the isolated muonic hydrogen atom and the average value
of the Coulomb repulsion of the nuclei. At $R=24$, this potential
agrees with the polarization potential with two percent accuracy.
As $R$ decreases further, the quasicrossings occur closer and
closer to the barrier top, the quasicrossing regions become
broader, and the statement that the muon is localized near one of
the nuclei loses meaning. In our case, the quasicrossings at
$R=4-6\mbox{ and }7-8$ occur near the barrier top.
%
%
%
%
\newpage
\begin{figure}[h]
\begin{center}
\includegraphics[scale=0.75]{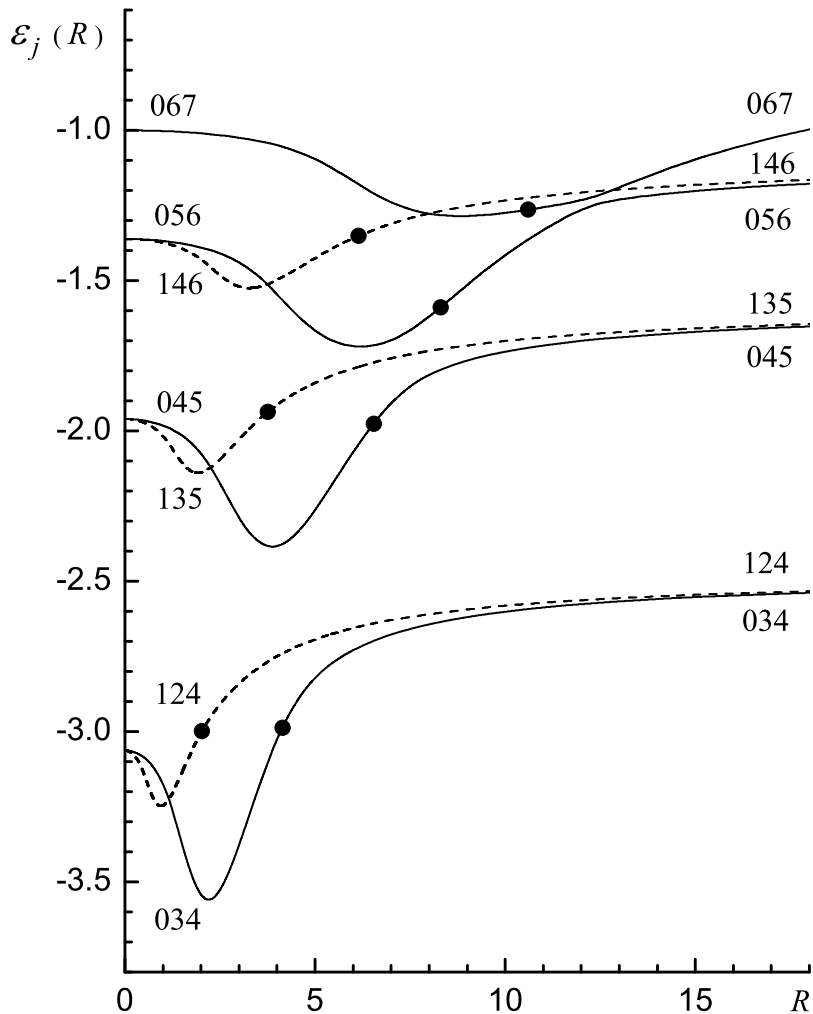}
\caption{Eigenvalues ${\varepsilon_j}(R)$ of the two--center
Coulomb problem versus the interatomic distance $R$ for some
right--center states with $m=0$. All the values are given in
muon--atom units. Each curve is marked with the values of the
parabolic quantum numbers $n'_1$, $n'_2$, and $n'$. The solid and
dashed curves correspond respectively to states with $n'_1=0\mbox{
and }1$. The black circle on each curve indicates the value of $R$
at which the respective state crosses the top of the potential
barrier separating the Coulomb wells of the two--center problem.}
\end{center}
\end{figure}
%

According to the generally accepted viewpoint, the muon transfer
is due to not too distant quasicrossings occurring near the
barrier top. For example, in Refs.~\cite{ger,sav}, the muon
transfer to oxygen was considered on the basis of an analysis of
terms of the two--center Coulomb problem with the charges of 1 and
8. It was found that the main contribution to the transfer rate
came from quasicrossings for which the distance between the
Coulomb centers lies in the interval of $8-9$. By following these
ideas, four right--center states were taken into account in the
expansion of the three--body wave function. Their quantum numbers
are $m=n'_1=0$ and $n'_2=3-6$. Moreover, according to
Ref.~\cite{ger}, states with $m=0$ and $n'_1=1$ can play some
role. Therefore, three such states with the quantum number
$n'_2=2-4$ were also included into the calculation. For these
states, the dependences of the eigenvalues ${\varepsilon_j}$ on
the distance $R$ are also shown in Fig.~2. Their two--center wave
functions have one node in the variable $\xi$; the number of nodes
in the variable $\eta$ is equal to $n'_2$. Thus, the basis used
was composed of seven right--center states with $m=0$. For
convenience, their quantum numbers are presented in
Table~\ref{tab4}.
%
%
%
%
\begin{table}[h]
\caption{Parabolic quantum numbers $n'_1$ and $n'_2$ of the
right--center states with $m=0$ in the limit of $R\to\infty$.
$n'=n'_1+n'_2+m+1$ is the principle quantum number. In the region
of $13<R<24$, the state marked by an asterisk describes the muonic
hydrogen atom in the Coulomb field of the oxygen nucleus. At
$R\approx 24$, it corresponds to the entrance reaction channel.
The remaining states describe the transfer channel.}
\bigskip
\begin{tabular}{|c|c|c|}
\hline
$\quad n'_1\quad$&$\quad n'_2\quad$&$\quad n'\quad$\\
\hline
\hspace{-2mm}*\,0&6&7\\
0&5&6\\
0&4&5\\
0&3&4\\
\hline
1&4&6\\
1&3&5\\
1&2&4\\
\hline
\end{tabular}
\label{tab4}
\end{table}
%

When using the above two--center basis, there is a set of seven
coupled equations for radial functions in the region on the left
of the quasicrossing at $R\approx 24.3$. In the region of
$13<R<24$, the right--center state with the quantum numbers
$m=n'_1=0$ and $n'_2=6$ describes the muonic hydrogen atom in the 
field of the oxygen nucleus. The matrix elements coupling the
equation for the radial function of this state with the remaining
equations decrease exponentially with increasing $R$. Therefore,
at $R\approx 24$ this equation splits off from the others and
corresponds to the entrance reaction channel. As already
mentioned, the adiabatic potential in this equation agrees with
the polarization potential with two percent accuracy. At $R\geq
24$, the equation~(\ref{xi0}) with the polarization potential
$U_p(R)$ was used to describe the entrance channel, i.e. the deep
subbarrier quasicrossings lying at $R\approx 24.3\mbox{ and }66$
were ignored. In this region of $R$, the muon transfer channel was
described by six radial equations for the remaining states
presented in Table~\ref{tab4}.

The above consideration corresponds to the muon transfer to a bare
oxygen nucleus. In fact, the muonic hydrogen atom collides with a
molecule. In view of the complexity of this process, let us
consider a collision of the muonic atom with a free oxygen atom.
Even in this simplified case, an analysis of dynamics of the
electron shell in the collision is a complicated problem. Indeed,
the energy released in the muon transfer reaction is a few keV,
and it is more than sufficient for electron excitations. The
simplest approximation is to ignore the excitations and to assume
that the electron shell remains in the ground state during the
collision. In this case, the role of the electron shell reduces to
screening the Coulomb interaction between heavy particles in
reaction channels. It is natural to expect that, because of low
collision energies, the screening is most significant in the
entrance channel. In the present study, the electron screening was
taken into account in Eq.~(\ref{xi0}), which describes the
entrance channel at $R\geq 24$. Instead of the polarization
potential $U_p(R)$, a new spherically symmetric potential $U_e(R)$
was used. It was constructed by the method proposed in
Ref.~\cite{krav}. This potential can be written as the sum of two
terms:
\begin{equation}
\label{Ue}
U_e(R)=U_s(R)+U_f(R)\,.
\end{equation}
The first term is the screened polarization potential:
\begin{equation}
\label{Us}
U_s(R)=-\frac{\beta Z_a^{\,2}(R)}{2R^{\,4}}\,,\quad
Z_a(R)=Z-Z_e(R)\,.
\end{equation}
$Z_e(R)$ is the absolute value of the electron charge in a sphere
of radius $R$ with the center at the nucleus of the atom, $Z_a(R)$
is the total charge of the atom in this sphere. The second term is
due to a finite size of the muonic hydrogen atom. It can be
considered as a contact interaction of the muonic atom with the
electron shell:
\begin{equation}
\label{Uf}
U_f(R)=\frac{2\pi}{3}\,<r_{\mu\rm H}^{\,2}>\rho_e(R)\,,
\end{equation}
where $<r_{\mu\rm H}^{\,2}>$ is the mean square of the
ground--state charge radius of the muonic atom with respect to its
center of mass:
\begin{equation}
\label{r2}
<r_{\mu\rm H}^{\,2}>=-\frac{3}{m_{\mu\rm H}}\left(1-\frac{1}{M_{\rm
H}}\right)\,.
\end{equation}
The square of the charge radius is negative because it is mainly
contributed by the negatively charged muon. The function
$\rho_e(R)$ is the absolute value of the electron density at the
distance $R$ from the $Z$ nucleus. It is normalized by the
condition
\begin{equation}
\label{roe}
4\pi\int\limits_0^\infty\rho_e(R)\,R^{\,2} dR=Z\,.
\end{equation}

The values of the potentials $U_s(R)$ and $U_f(R)$ are presented
in Table~\ref{tab5} for a number of interatomic distances. The
electron density $\rho_e(R)$ and the charge $Z_e(R)$ were
calculated with the aid of analytic one--electron wave functions
obtained by the Hartree-Fock-Roothaan method~\cite{clem}. Both the
potentials $U_s(R)$ and $U_f(R)$ are attractive and decrease
exponentially with increasing $R$. Since $U_s(R)$ is proportional
to the square of the atomic charge $Z_a(R)$ and involves the
additional factor $R^{\,-4}$, it decreases faster. As a result,
this potential is significant only at distances $R$ that do not
exceed the electron Bohr radius (about 200 muon--atom units). For
example, at $R=24$ the potential $U_s=-2.92$~eV, and it is one
order of magnitude greater than the term $U_f$\,. Since the
electron $K$--shell of the oxygen atom has approximately the same
radius, the screening in the potential $U_s$ is already
noticeable: the atomic charge $Z_a=7.45$\,. At $R\approx 105$, the
potentials $U_s$ and $U_f$ become equal each other, and their sum
is about $-8.5\times 10^{\,-3}$~eV. This value is of the order of
thermal energies at a temperature of 100~K. At $R=200$, the term
$U_s$ is about 9\,\% of the potential $U_f$\,. In this case,
$U_e=1.69\times 10^{\,-3}$~eV, and it corresponds to thermal
energies at a temperature of 20~K. One more fact associated with
the potential $U_f$ should be noted. The electron screening
weakens the polarization attraction, so that the inequality
$|\,U_s\,|<|\,U_p\,|$ holds at all values of $R$\,. The addition
of the term $U_f$ leads to the new potential $U_e$ becoming
greater in absolute value than $U_p$ at $R>115$:
$|\,U_e\,|>|\,U_p\,|$\,. Since $U_e$ decreases exponentially,
while $U_p$ follows the power law $1/R^{\,4}$, the opposite
inequality holds at $R>600$\,. The potentials are already very
small at this point. They are about $-8\times 10^{\,-6}$~eV, which
corresponds to a temperature of about 0.1~K.
%
%
%
%
\begin{table}[h]
\caption{Atomic charge $Z_a$ and potential energies of interaction
between the muonic hydrogen atom and the oxygen atom versus the
interatomic distance $R$\,. The values of $R$ and $Z_a$ are given
in muon--atom units, the potential energies are given in eV. The
orders of values are indicated parenthetically.}
\bigskip
\begin{tabular}{|c|c|l|l|l|l|}
\hline
$\hspace{4mm}R\hspace{4mm}$&$\hspace{4mm}Z_a\hspace{4mm}$&$\hspace{8mm}U_p$&$\hspace{8mm}U_s$&$\hspace{8mm}U_f$&$\hspace{8mm}U_e$\\
\hline
24&7.45&$\;-3.36$&$\;-2.92$&$\;-1.97\,(-1)\;$&$\;-3.11$\\
30&7.20&$\;-1.38$&$\;-1.12$&$\;-1.26\,(-1)\;$&$\;-1.24$\\
40&6.82&$\;-4.36\,(-1)\;$&$\;-3.17\,(-1)\;$&$\;-6.09\,(-2)\;$&$\;-3.77\,(-1)\;$\\
50&6.50&$\;-1.79\,(-1)\;$&$\;-1.18\,(-1)\;$&$\;-3.05\,(-2)\;$&$\;-1.49\,(-1)\;$\\
75&5.99&$\;-3.53\,(-2)\;$&$\;-1.98\,(-2)\;$&$\;-7.89\,(-3)\;$&$\;-2.77\,(-2)\;$\\
100&5.61&$\;-1.12\,(-2)\;$&$\;-5.49\,(-3)\;$&$\;-4.41\,(-3)\;$&$\;-9.89\,(-3)\;$\\
125&5.17&$\;-4.57\,(-3)\;$&$\;-1.91\,(-3)\;$&$\;-3.45\,(-3)\;$&$\;-5.36\,(-3)\;$\\
150&4.65&$\;-2.20\,(-3)\;$&$\;-7.44\,(-4)\;$&$\;-2.74\,(-3)\;$&$\;-3.48\,(-3)\;$\\
175&4.08&$\;-1.19\,(-3)\;$&$\;-3.09\,(-4)\;$&$\;-2.10\,(-3)\;$&$\;-2.41\,(-3)\;$\\
200&3.50&$\;-6.97\,(-4)\;$&$\;-1.34\,(-4)\;$&$\;-1.56\,(-3)\;$&$\;-1.69\,(-3)\;$\\
250&2.47&$\;-2.86\,(-4)\;$&$\;-2.73\,(-5)\;$&$\;-8.19\,(-4)\;$&$\;-8.46\,(-4)\;$\\
300&1.67&$\;-1.38\,(-4)\;$&$\;-6.03\,(-6)\;$&$\;-4.19\,(-4)\;$&$\;-4.25\,(-4)\;$\\
400&0.715&$\;-4.36\,(-5)\;$&$\;-3.48\,(-7)\;$&$\;-1.09\,(-4)\;$&$\;-1.10\,(-4)\;$\\
500&0.291&$\;-1.79\,(-5)\;$&$\;-2.35\,(-8)\;$&$\;-2.95\,(-5)\;$&$\;-2.96\,(-5)\;$\\
600&0.115&$\;-8.61\,(-6)\;$&$\;-1.79\,(-9)\;$&$\;-8.27\,(-6)\;$&$\;-8.27\,(-6)\;$\\
700&0.0455&$\;-4.65\,(-6)\;$&$\;-1.50\,(-10)\;$&$\;-2.40\,(-6)\;$&$\;-2.40\,(-6)\;$\\
\hline
\end{tabular}
\label{tab5}
\end{table}
%

In order to clarify the role of the electron screening, the
calculations of the reaction rate were performed for three
versions A, B, and C. They differ in the potential in
Eq.~(\ref{xi0}), which asymptotically describes the entrance
channel.
\begin{enumerate}
\item[A)] The electron screening was fully ignored. This
corresponds to the muon transfer to a bare oxygen nucleus. In this
case, the unscreened polarization potential $U_p(R)$ was used in
Eq.~(\ref{xi0}).
\item[B)] The potential $U_p(R)$ was replaced by the screened
polarization potential $U_s(R)$, i.e. the screening of the nuclear
charge by atomic electrons was taken into account. Such a way was
also applied in Refs.~\cite{sav,lin}.
\item[C)] The potential $U_e(R)=U_s(R)+U_f(R)$ was used in
Eq.~(\ref{xi0}). In this case, the contact interaction of the
muonic hydrogen atom with the electron shell of oxygen was added
to the screened polarization potential. The version C is the most
realistic because it takes into account the effect of atomic
electrons to a greater extent than the versions A and B.
\end{enumerate}

In all the versions, the effective potential appearing in
Eq.~(\ref{xi0}) features a barrier at nonzero values of the
orbital angular momentum $J$. The position of the barrier top
$R_b$ and its height $U_b$ are given in Table~\ref{tab6} for
$J\leq 4$. At these values of $J$, the barrier top lies either in
the region of $R\geq 24$, where the entrance channel is described
by Eq.~(\ref{xi0}) alone, or near this region on the left from it.
At low collision energies $E\ll U_b$\,, the barrier prevents the
penetration of the respective partial wave into the
term--interaction region, and the contribution of this wave to the
muon transfer cross section is small. As the collision energy
grows, the partial cross section of the muon transfer increases
and at $E\sim U_b$ becomes commensurate with the contributions of
waves with lower orbital angular momenta.
%
%
%
%
\begin{table}[h]
\caption{Position $R_b$ of the top of the potential barrier and
its height $U_b$ for several values of the orbital angular
momentum $J$ according to calculations in the versions A, B, and
C. The values of $R_b$ are given in muon--atom units, the values
of $U_b$ are given in eV.}
\bigskip
\begin{tabular}{|c|c|c|c|c|c|c|c|c|}
\hline
$\quad\mbox{Version of}\quad$ & \multicolumn{2}{c|}{$J=1$} & \multicolumn{2}{c|}{$J=2$} & \multicolumn{2}{c|}{$J=3$} & \multicolumn{2}{c|}{$J=4$}\\
\cline{2-9}
$\quad\mbox{calculation}\quad$ & $\quad R_b\quad$ & $\quad U_b\quad$ & $\quad R_b\quad$ & $\quad U_b\quad$ & $\quad R_b\quad$ & $\quad U_b\quad$ & $\quad R_b\quad$ & $\quad U_b\quad$\\
\hline
  A     & 60.5 & 0.0832 & 34.9 & 0.749 & 24.7 & 3.00 & 19.1 & 8.32\\
  B     & 51.4 & 0.126  & 32.4 & 0.943 & 23.8 & 3.43 & 18.7 & 9.07\\
  C     & 55.9 & 0.102  & 33.5 & 0.841 & 24.1 & 3.24 & 18.8 & 8.77\\
\hline
\end{tabular}
\label{tab6}
\end{table}
%
%
\section{\rm Results of the calculation and conclusions}
\label{results}
%
%
The results of the present calculation of the muon transfer rate
$q(E)$ are given in Figs.~3 and 4, and also in
Appendix~\ref{appA}. At low collision energies, the $s$--wave
contribution predominates, the transfer cross section is
proportional to $1/v$\,, and the transfer rate is nearly constant.
At $E>0.01$~eV, the contribution of the $p$--wave grows fast and
becomes decisive. As a result, pronounced resonance maxima arise
on the curves of the dependence $q(E)$. Their position and the
maximum transfer rate obtained within the versions A, B, and C are
given in Table~\ref{tab7}. The position and the shape of the
maxima depend substantially on the electron screening. The
sharpest maximum is obtained in the version A (the muon tranfer to
a bare oxygen nucleus). In this case, the maximum value of the
transfer rate is reached at a collision energy which is somewhat
less than the height of the potental barrier in the $p$--wave. In
the version B, taking into account the electron screening in the
polarization potential reduces the attraction in the entrance
reaction channel. As a result, the maximum shifts toward higher
energies, and its height decreases. In the most realistic version
C, the additional attraction caused by the contact interaction of
the muonic hydrogen atom with the electron shell of oxygen shifts
the maximum back to lower energies and increases its height. In
the versions B and C, the maximum is reached at collision energies
which are somewhat greater than the height of the potential
barrier in the $p$--wave. As the collision energy grows further,
the electron screening becomes less significant. As a result, the
curves obtained in all the three versions of the calculation are
almost coincident at $E>1$~eV. In this region, there is one more
maximum at $E\approx 2.4$~eV. The transfer rate is about $5\times
10^{\,10}\mbox{ s}^{-1}$ at this energy. This maximum is due to a
resonance behaviour of the partial contribution of the $g$--wave
(Fig.~4). It is interesting to note that the height of the
potential barrier in this wave is about 9~eV (Table~\ref{tab6}),
so that this resonance is deep--subbarrier.
%
%
%
%
\begin{figure}[h]
\begin{center}
\includegraphics[scale=0.75]{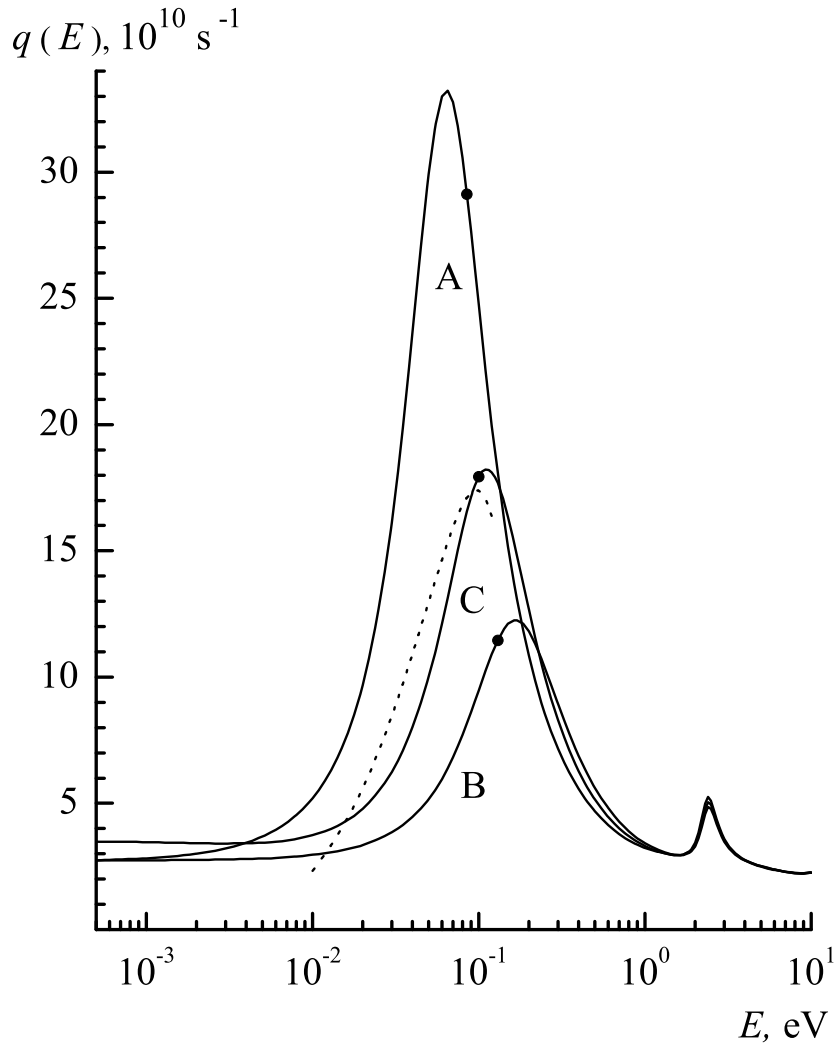}
\caption{Muon transfer rate $q(E)$ versus the collision energy
$E$. The solid curves reprsent the results obtained within the
versions A, B, and C. The black circle on each curve indicates the
value of $E$ which is equal to the height of the barrier in the
effective potential energy for the $p$--wave (Table~\ref{tab6}).
The dotted curve represents the experimental dependence of $q(E)$.
It was calculated by the formula~(\ref{rate2}) with the central
values of the coefficients from Table~\ref{tab2}.}
\end{center}
\end{figure}
%
%
%
%
\newpage
\begin{figure}[h]
\begin{center}
\includegraphics[scale=0.75]{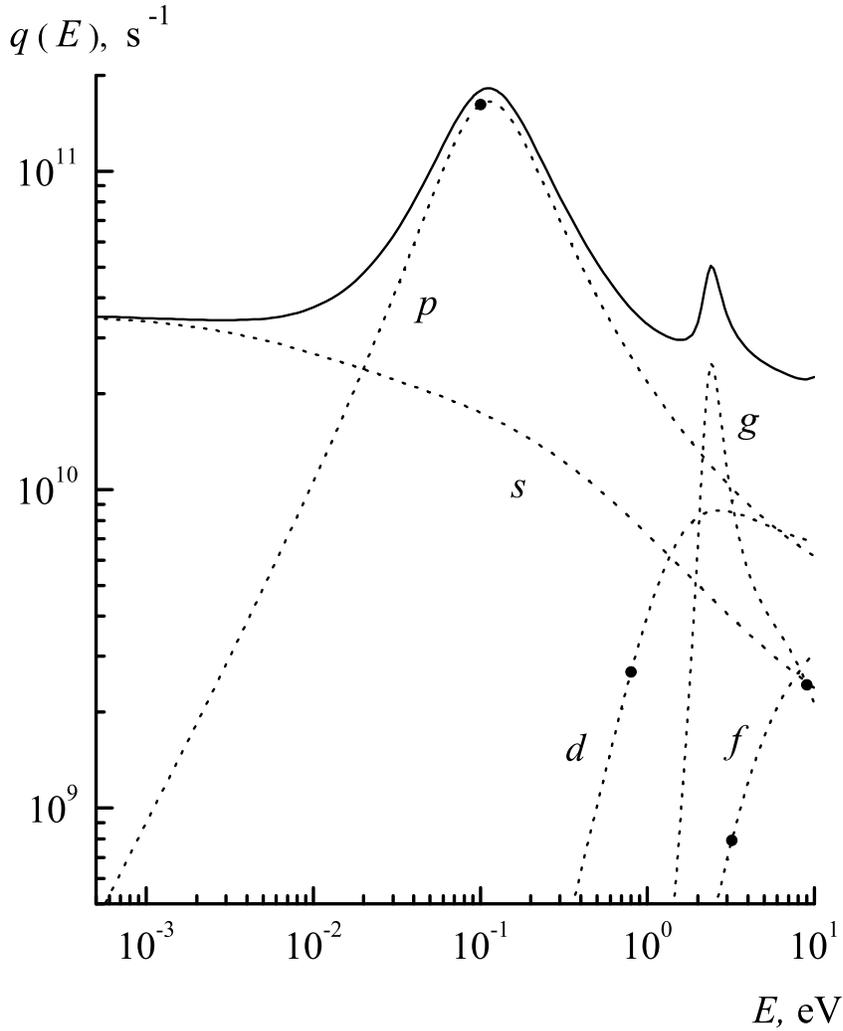}
\caption{Partial muon tranfer rates versus the collision energy
$E$. The curves were calculated within the version C. The solid
curve represents the total transfer rate $q(E)$, the dotted curves
are the contributions of the partial waves from $s$ to $g$. The
black circle on each dotted curve indicates the value of $E$ which
is equal to the height of the barrier in the effective potential
energy for the respective partial wave (Table~\ref{tab6}).}
\end{center}
\end{figure}
%
%
%
%
\newpage
\begin{table}[h]
\caption{Position $E_m$ of the resonance maxima in the energy
dependence of the muon transfer rate at collision energies
$E<1$~eV. $q_m$ is the maximum transfer rate. The last row is the
results of the approximation proposed in Ref.~\cite{adam4} (the
formula~(\ref{rate2}) with the central values of the coefficients
from Table~\ref{tab2}).}
\bigskip
\begin{tabular}{|c|c|c|}
\hline
$\quad\mbox{Version of}\quad$ & $\quad E_m,\,\mbox{eV}\quad$ & $\quad q_m,\,10^{\,11}\,\mbox{s}^{-1}\quad$\\
$\quad\mbox{calculation}\quad$ &  & \\
\hline
    A        & 0.0640 & 3.32 \\
    B        & 0.167  & 1.23 \\
    C        & 0.112  & 1.82 \\
\hline
\cite{adam4} & 0.0972 & 1.74 \\
\hline
\end{tabular}
\label{tab7}
\end{table}
%

The experimental curve of the dependence $q(E)$ is also shown in
Fig.~3. It was obtained in Ref.~\cite{adam4} as a result of the
processing of data on the temperature dependence of the transer
rate. This curve was calculated by the formula~(\ref{rate2}) with
the central values of the coefficients from Table~\ref{tab2}. The
position and the height of the maximum of this curve are given in
Table~\ref{tab7}. In the region of $E>0.015$~eV, the curve
calculated within the version C agrees well with the experimental
curve. The calculated and experimental values of the position of
the maximum and its height are also close. At lower collision
energies, the agreement becomes worse. It is possible that this is
due to molecular effects. In this context, two circumstances
should be noted.
\begin{enumerate}
\item The logic of the present calculation is that the muon
transfer occurs at interatomic distances $R$ not exceeding 24
muon--atom units. This value is about one tenth of the electron
Bohr radius, and it is noticeably less than dimensions of the
oxygen molecule. At greater values of $R$, the muonic hydrogen
atom moves in the spherically symmetric field of the free oxygen
atom. Actually, the molecular field is much more complicated. In
particular, it is not spherically symmetric. It is natural to
expect that this fact may be significant at low collision
energies.
\item The nuclei of the oxygen molecule take part in internal
vibrational--rotational motion. In particular, there are
zero--point vibrations at any temperature. Their consideration may
be also important at low collision energies.
\end{enumerate}

Let us now consider the temperature dependence of the rate
$\lambda(T)$ of the muon transfer from thermalized muonic hydrogen
atoms. For a collision of the muonic atom with a free oxygen atom,
this quantity is obtained by averaging the rate $q(E)$ over the
Maxwellian distribution of relative velocities in the entrance
reaction channel. The results are presented in Fig.~5, in
Table~\ref{tab8}, and also in Appendix~\ref{appB}. In all the
three versions of the calculation, the transfer rate increases
monotonically with temperature. This results from the growth of
the rate $q(E)$ at thermal collision energies. The best agreement
with experimental data is obtained in the version C. At $T>150$~K,
the calculated curve lies somewhat lower than experimental points.
Its deviations from these points do not exceed 15~\%. At lower
temperatures, the agreement is poorer. The calculated values
become greater than the experimental ones, and at $T\leq 104$~K
exceed the latter by 30 -- 40~\%. This fact may also be an
indication of the need to take into account molecular effects.
%
%
%
%
\newpage
\begin{figure}[h]
\begin{center}
\includegraphics[scale=0.75]{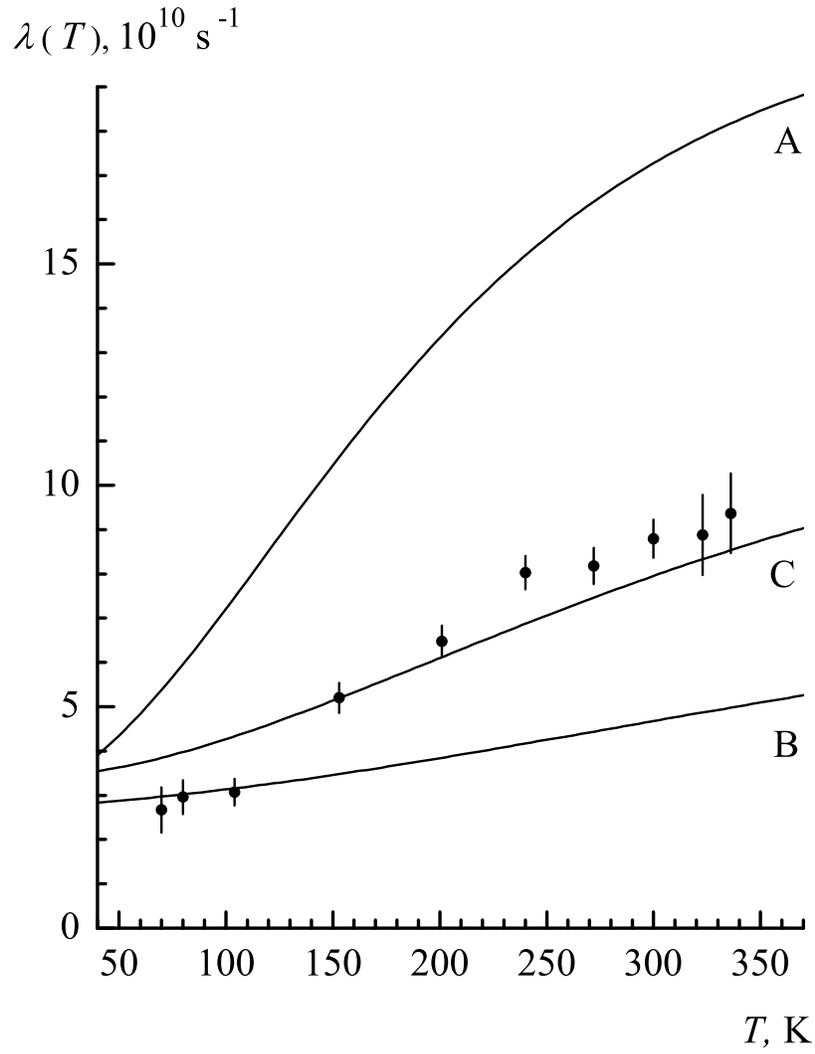}
\caption{Rate $\lambda(T)$ of the muon transfer from thermalized
muonic hydrogen atoms versus temperature. The solid curves
reprsent the results obtained within the versions A, B, and C. The
black circles are the experimental values obtained in
Refs.~\cite{adam5,adam6} (Table~\ref{tab8}).}
\end{center}
\end{figure}
%
%
%
%
\newpage
\begin{table}[h]
\caption{Experimental $\lambda_{\,\rm exp}$ and calculated
$\lambda_{\,\rm th}$ values of the rate $\lambda(T)$ of the muon
transfer from thermalized muonic hydrogen atoms to oxygen at
various temperatures. All the rates are given in unuts of
$10^{\,10}\,\mbox{s}^{-1}$. The experimental values were taken
from Refs.~\cite{adam5,adam6}. The errors were calculated by the
formula $\sigma=\sqrt{\,\sigma_1^{\,2}+\sigma_2^{\,2}}$\,, where
$\sigma_1$ and $\sigma_2$ are given in Table~\ref{tab1}. The
values of $\lambda_{\,\rm th}$ were calculated within the versions
A, B, and C.}
\bigskip
\begin{tabular}{|c|c|c|c|c|}
\hline
$\quad T,~{\rm K}\quad$&$\lambda_{\,\rm exp}$&\multicolumn{3}{c|}{$\lambda_{\,\rm th}$}\\
\cline{3-5}
  &\cite{adam5,adam6}& A & B & C\\
\hline
70&$\quad 2.67\pm 0.51\quad$&$\quad 5.38\quad$&$\quad 2.97\quad$&$\quad 3.84\quad$\\
80&$2.96\pm 0.38$&$5.96$&$3.02$&$3.97$\\
104&$3.07\pm 0.30$&$7.47$&$3.16$&$4.33$\\
153&$5.20\pm 0.34$&10.6&3.48&5.19\\
201&$6.48\pm 0.35$&13.4&3.85&6.12\\
240&$8.03\pm 0.38$&15.2&4.17&6.87\\
272&$8.18\pm 0.41$&16.4&4.44&7.47\\
300&$8.79\pm 0.43$&17.3&4.68&7.95\\
323&$8.88\pm 0.91$&17.9&4.87&8.33\\
336&$9.37\pm 0.90$&18.2&4.98&8.54\\
\hline
\end{tabular}
\label{tab8}
\end{table}
%

In conclusion, let us review some results.
\begin{enumerate}
\item In the collision energy region of $0.015<E<0.1$~eV, the
calculated values of the transfer rate agree well with the results
obtained by the processing of experimental data. In particular,
the calculation predicts the existence of a $p$--wave resonance at
an energy of $E\approx 0.1$~eV. This result seems to be important
in the context of the planned laser experiment on precise
measurements of the hyperfine splitting energy of the $1s$ state
of muonic hydrogen.
\item At temperatures of $T\geq 150$~K, the calculated values of
the rate of the muon transfer from thermalized muonic atoms are
also in good agreement with experimental values.
\item The results of the calculation are quite sensitive to the
electron screening. Good agreement with experimental data is
obtained only with its proper consideration.
\item As the collision energy decreases to 0.01~eV and the
temperature decreases to 100~K, the agreement between callculated
and experimental values becomes poorer. A possible reason may lie
in molecular effects.
\end{enumerate}
%
%
%
%
\newpage
\appendix
\section{}
\label{appA}
Tables given below represent numerical values of the muon transfer
rate $q(E)$ calculated for a number of collision energies $E$
within the versions A, B, and C.
\begin{table}[h]
\begin{tabular}{|c|c|c|c|}
\hline
$\quad E,\,{\rm eV}\quad$&\multicolumn{3}{c|}{$\quad q(E),\,10^{10}\mbox{ s}^{-1}\quad$}\\
\cline{2-4}
  & A & B & C \\
\hline
$1\times 10^{-4}$&$\quad 2.70\quad$&$\quad 2.74\quad$&$\quad 3.54\quad$\\
$2\times 10^{-4}$&$      2.71     $&$      2.74     $&$      3.53     $\\
$4\times 10^{-4}$&$      2.73     $&$      2.73     $&$      3.51     $\\
$6\times 10^{-4}$&$      2.76     $&$      2.74     $&$      3.49     $\\
$8\times 10^{-4}$&$      2.79     $&$      2.74     $&$      3.47     $\\
\hline
$1\times 10^{-3}$&$      2.82     $&$      2.74     $&$      3.46     $\\
$2\times 10^{-3}$&$      3.00     $&$      2.75     $&$      3.42     $\\
$4\times 10^{-3}$&$      3.44     $&$      2.80     $&$      3.41     $\\
$6\times 10^{-3}$&$      3.95     $&$      2.85     $&$      3.48     $\\
$8\times 10^{-3}$&$      4.54     $&$      2.90     $&$      3.59     $\\
\hline
$0.01 $          &$      5.20     $&$      2.96     $&$      3.74     $\\
$0.015$          &$      7.19     $&$      3.14     $&$      4.21     $\\
$0.02 $          &$      9.68     $&$      3.35     $&$      4.79     $\\
$0.025$          &$      12.7     $&$      3.59     $&$      5.47     $\\
$0.03 $          &$      16.1     $&$      3.85     $&$      6.24     $\\
$0.035$          &$      19.9     $&$      4.14     $&$      7.09     $\\
$0.04 $          &$      23.6     $&$      4.46     $&$      8.01     $\\
$0.045$          &$      27.1     $&$      4.80     $&$      8.99     $\\
$0.05 $          &$      29.9     $&$      5.17     $&$      10.0     $\\
$0.055$          &$      31.9     $&$      5.56     $&$      11.1     $\\
$0.06 $          &$      33.0     $&$      5.96     $&$      12.1     $\\
$0.065$          &$      33.2     $&$      6.39     $&$      13.1     $\\
$0.07 $          &$      32.8     $&$      6.82     $&$      14.1     $\\
$0.075$          &$      31.8     $&$      7.27     $&$      15.0     $\\
$0.08 $          &$      30.6     $&$      7.72     $&$      15.9     $\\
$0.085$          &$      29.1     $&$      8.17     $&$      16.6     $\\
$0.09 $          &$      27.6     $&$      8.61     $&$      17.1     $\\
$0.093$          &$       -       $&$       -       $&$      17.4     $\\
$0.095$          &$      26.2     $&$      9.05     $&$      17.6     $\\
\hline
\end{tabular}
\end{table}
%
%
\newpage
\begin{table}[h]
\begin{tabular}{|c|c|c|c|}
\hline
$\quad E,\,{\rm eV}\quad$&\multicolumn{3}{c|}{$\quad q(E),\,10^{10}\mbox{ s}^{-1}\quad$}\\
\cline{2-4}
  & A & B & C \\
\hline
$0.1  $          &$\quad 24.8\quad$&$\quad 9.47\quad$&$\quad 17.9\quad$\\
$0.105$          &$       -       $&$       -       $&$      18.1     $\\
$0.11 $          &$      22.2     $&$      10.2     $&$      18.2     $\\
$0.115$          &$       -       $&$       -       $&$      18.2     $\\
$0.12 $          &$      20.0     $&$      10.9     $&$      18.1     $\\
$0.125$          &$       -       $&$       -       $&$      17.9     $\\
$0.13 $          &$      18.1     $&$      11.5     $&$      17.7     $\\
$0.135$          &$       -       $&$       -       $&$      17.4     $\\
$0.14 $          &$       -       $&$       -       $&$      17.1     $\\
$0.145$          &$       -       $&$       -       $&$      16.7     $\\
$0.15 $          &$      15.2     $&$      12.1     $&$      16.3     $\\
$0.16 $          &$       -       $&$       -       $&$      15.6     $\\
$0.17 $          &$      13.1     $&$      12.3     $&$      14.8     $\\
$0.18 $          &$       -       $&$       -       $&$      14.0     $\\
$0.2  $          &$      10.9     $&$      11.8     $&$      12.7     $\\
$0.22 $          &$       -       $&$       -       $&$      11.5     $\\
$0.25 $          &$      8.57     $&$      10.4     $&$      10.0     $\\
$0.3  $          &$      7.17     $&$      8.94     $&$      8.30     $\\
$0.35 $          &$      6.24     $&$      7.78     $&$      7.12     $\\
$0.4  $          &$      5.57     $&$      6.88     $&$      6.27     $\\
$0.45 $          &$      5.07     $&$      6.17     $&$      5.64     $\\
$0.5  $          &$      4.69     $&$      5.62     $&$      5.15     $\\
$0.6  $          &$      4.14     $&$      4.82     $&$      4.47     $\\
$0.7  $          &$      3.78     $&$      4.28     $&$      4.01     $\\
$0.8  $          &$      3.53     $&$      3.89     $&$      3.70     $\\
$0.9  $          &$      3.36     $&$      3.62     $&$      3.47     $\\
\hline
\end{tabular}
\end{table}
%
%
\newpage
\begin{table}[h]
\begin{tabular}{|c|c|c|c|}
\hline
$\quad E,\,{\rm eV}\quad$&\multicolumn{3}{c|}{$\quad q(E),\,10^{10}\mbox{ s}^{-1}\quad$}\\
\cline{2-4}
  & A & B & C \\
\hline
$1.0  $          &$\quad 3.24\quad$&$\quad 3.42\quad$&$\quad 3.31\quad$\\
$1.2  $          &$      3.07     $&$      3.15     $&$      3.10     $\\
$1.5  $          &$      2.96     $&$      2.97     $&$      2.96     $\\
$1.7  $          &$      2.97     $&$      2.96     $&$      2.96     $\\
$1.9  $          &$      3.18     $&$      3.10     $&$      3.13     $\\
$2.0  $          &$      3.43     $&$      3.29     $&$      3.35     $\\
$2.1  $          &$      3.85     $&$      3.60     $&$      3.71     $\\
$2.2  $          &$      4.42     $&$      4.07     $&$      4.23     $\\
$2.3  $          &$      4.99     $&$      4.57     $&$      4.76     $\\
$2.4  $          &$      5.25     $&$      4.85     $&$      5.04     $\\
$2.5  $          &$      5.10     $&$      4.79     $&$      4.94     $\\
$2.6  $          &$      4.75     $&$      4.51     $&$      4.63     $\\
$2.7  $          &$      4.37     $&$      4.18     $&$      4.28     $\\
$2.8  $          &$      4.05     $&$      3.90     $&$      3.97     $\\
$2.9  $          &$      3.79     $&$      3.67     $&$      3.73     $\\
$3.0  $          &$      3.59     $&$      3.49     $&$      3.54     $\\
$3.2  $          &$      3.30     $&$      3.22     $&$      3.26     $\\
$3.5  $          &$      3.02     $&$      2.97     $&$      3.00     $\\
$4.0  $          &$      2.77     $&$      2.74     $&$      2.75     $\\
$5.0  $          &$      2.51     $&$      2.50     $&$      2.51     $\\
$6.0  $          &$      2.38     $&$      2.37     $&$      2.37     $\\
$7.0  $          &$      2.29     $&$      2.29     $&$      2.29     $\\
$8.0  $          &$      2.24     $&$      2.24     $&$      2.24     $\\
$9.0  $          &$      2.23     $&$      2.22     $&$      2.22     $\\
$10.0 $          &$      2.26     $&$      2.25     $&$      2.25     $\\
\hline
\end{tabular}
\end{table}
%
%
%
%
\newpage
\section{}
\label{appB}
Tables given below represent numerical values of the rate
$\lambda(T)$ of the muon transfer from thermalized muonic hydrogen
atoms for a number of temperatures $T$. These values were
calculated within the versions A, B, and C.
\begin{table}[h]
\begin{tabular}{|c|c|c|c|}
\hline
$\quad T,\,{\rm K}\quad$&\multicolumn{3}{c|}{$\quad \lambda(T),\,10^{10}\mbox{ s}^{-1}\quad$}\\
\cline{2-4}
  & A & B & C \\
\hline
$20  $&$\quad 3.17\quad$&$\quad 2.77\quad$&$\quad 3.45\quad$\\
$25  $&$      3.33     $&$      2.79     $&$      3.46     $\\
$30  $&$      3.51     $&$      2.80     $&$      3.48     $\\
$35  $&$      3.70     $&$      2.82     $&$      3.51     $\\
$40  $&$      3.90     $&$      2.84     $&$      3.55     $\\
$45  $&$      4.11     $&$      2.86     $&$      3.58     $\\
$50  $&$      4.34     $&$      2.88     $&$      3.63     $\\
$55  $&$      4.59     $&$      2.90     $&$      3.68     $\\
$60  $&$      4.84     $&$      2.92     $&$      3.73     $\\
$65  $&$      5.11     $&$      2.95     $&$      3.79     $\\
$70  $&$      5.38     $&$      2.97     $&$      3.84     $\\
$75  $&$      5.67     $&$      2.99     $&$      3.91     $\\
$80  $&$      5.96     $&$      3.02     $&$      3.97     $\\
$85  $&$      6.27     $&$      3.05     $&$      4.04     $\\
$90  $&$      6.58     $&$      3.07     $&$      4.11     $\\
$95  $&$      6.89     $&$      3.10     $&$      4.19     $\\
$100 $&$      7.21     $&$      3.13     $&$      4.27     $\\
\hline
$110 $&$      7.86     $&$      3.19     $&$      4.43     $\\
$120 $&$      8.52     $&$      3.25     $&$      4.59     $\\
$130 $&$      9.17     $&$      3.32     $&$      4.77     $\\
$140 $&$      9.82     $&$      3.39     $&$      4.95     $\\
$150 $&$      10.5     $&$      3.46     $&$      5.14     $\\
$160 $&$      11.1     $&$      3.53     $&$      5.33     $\\
$170 $&$      11.7     $&$      3.60     $&$      5.52     $\\
$180 $&$      12.2     $&$      3.68     $&$      5.71     $\\
$190 $&$      12.8     $&$      3.76     $&$      5.91     $\\
$200 $&$      13.3     $&$      3.84     $&$      6.10     $\\
\hline
\end{tabular}
\end{table}
%
%
%
\newpage
\begin{table}[h]
\begin{tabular}{|c|c|c|c|}
\hline
$\quad T,\,{\rm K}\quad$&\multicolumn{3}{c|}{$\quad \lambda(T),\,10^{10}\mbox{ s}^{-1}\quad$}\\
\cline{2-4}
  & A & B & C \\
\hline
$210 $&$\quad 13.8\quad$&$\quad 3.92\quad$&$\quad 6.30\quad$\\
$220 $&$      14.3     $&$      4.00     $&$      6.49     $\\
$230 $&$      14.8     $&$      4.08     $&$      6.68     $\\
$240 $&$      15.2     $&$      4.17     $&$      6.87     $\\
$250 $&$      15.6     $&$      4.25     $&$      7.06     $\\
$260 $&$      16.0     $&$      4.34     $&$      7.25     $\\
$270 $&$      16.3     $&$      4.42     $&$      7.43     $\\
$280 $&$      16.7     $&$      4.51     $&$      7.61     $\\
$290 $&$      17.0     $&$      4.59     $&$      7.78     $\\
$300 $&$      17.3     $&$      4.68     $&$      7.95     $\\
\hline
$310 $&$      17.5     $&$      4.76     $&$      8.12     $\\
$320 $&$      17.8     $&$      4.84     $&$      8.28     $\\
$330 $&$      18.0     $&$      4.93     $&$      8.44     $\\
$340 $&$      18.3     $&$      5.01     $&$      8.60     $\\
$350 $&$      18.5     $&$      5.09     $&$      8.75     $\\
$360 $&$      18.6     $&$      5.18     $&$      8.89     $\\
$370 $&$      18.8     $&$      5.26     $&$      9.03     $\\
$380 $&$      19.0     $&$      5.34     $&$      9.17     $\\
$390 $&$      19.1     $&$      5.42     $&$      9.30     $\\
$400 $&$      19.3     $&$      5.49     $&$      9.43     $\\
\hline
\end{tabular}
\end{table}
%
%
%
%
\newpage
%

%
%
\end {document}